\documentclass[a4paper,10pt]{article}

\usepackage{amssymb} 
\usepackage{latexsym}

\begin{document}
\LARGE
Yang-Mills Theory for Noncommutative Flows 
\begin{center}
Addendum
\end{center}
\vspace{1cm}
\begin{center}
\Large
Hiroshi TAKAI\\
\end{center}
\begin{center}
\Large
Department of Mathematics, \\
Tokyo Metropolitan University\\
\end{center}
\begin{center}
\Large
MinamiOhsawa,Hachiohji,Tokyo,JAPAN.\\
\end{center}

\vspace{2cm}
\begin{center}
\Large
Abstract \\
\end{center}
\vspace{2mm}

\Large
This supplementary manuscript is to describe an important nontrivial example, 
which appears in the matrix model of type IIB in the super string theory in 
order to apply a new duality for the moduli spaces of Yang-Mills connections 
on noncommutative vector bundles. Actually, the moduli space of the instanton 
bundle over noncommutative Euclidean 4-spaces with respect to the canonical 
action of space translations is computed precisely without using the ADHM-
construction.

\vspace{3cm}
\Large  
July,~2004

\newpage

\S1. Introduction  \quad In the manuscript [6], we have found a new duality for  the moduli spaces of Yang-Mills connections on noncommutative vector bundles 
with respect to noncommutative flows.  As we have also announced in [6] with a 
short proof that such a duality was also affirmatively shown for noncommutative multiflows. \\
\quad In this addendum, we apply it to compute the moduli space in the case of 
the instanton bundles on the noncommutative Euclidean 4-space with respect to 
the canonical space translations without using the ADHM construction (cf:[1],[2]). \\

\S2. Preliminaries  \quad  Let $(A,\mathbb{R}^n,\alpha),~(n\geq 1)$ be a $F^*$-
system, and $\Xi$ a finitely generated projective $\alpha$-invariant right $A$-module. As $\Xi=P(A^m)$ for a projection $P \in M_m(A)$ over $A~(m\geq 1)$, 
and let $\widehat{\Xi}=\widehat{P}(\widehat{A}^m)$ 
where $\widehat{A}=A\rtimes_{\alpha} \mathbb{R}^n$ 
and $\widehat{P}=P\times I \in M_m(\mathcal{M}(\widehat{A}))$ 
where $\mathcal{M}(\widehat{A})$ is the $F^*$-algebra consisting of all $A$-valued bounded $C^{\infty}$-functions on $\mathbb{R}^n$ with $\alpha$-twisted *-convolution. Then $\widehat{\Xi}$ is a finitely generated projective right $\widehat{A}$-module. In [6], we have shown the following theorem: \\

Theorem 1~([6]). ~Let $(A,\mathbb{R}^n,\alpha)$ be a $F^*$-dynamical system with a faithful $\alpha$-invariant continuous trace $\tau$, and $\Xi$ a finitely 
generated projective right $A$-module. Then there exist a dual $F^*$-dynamical 
system $(\widehat{A},\mathbb{R}^n,\overline{\alpha})$ with a faithful $\overline{\alpha}$-invariant trace $\widehat{\tau}$ and a finitely generated projective right $\widehat{A}$-module $\widehat{\Xi}$ with the property that the moduli 
space $\mathcal{M}^{(A,\mathbb{R}^n,\alpha,\tau)}(\Xi)$ of the Yang-Mills 
connections of $\Xi$ for $(A,\mathbb{R}^n,\alpha,\tau)$ is homeomorphic to the 
dual moduli space  $\mathcal{M}^{(\widehat{A},\mathbb{R}^n,\overline{\alpha},\widehat{\tau})}(\widehat{\Xi})$ of the Yang-Mills connections of $\widehat{\Xi}$ 
for $(\widehat{A},\mathbb{R}^n,\overline{\alpha},\widehat{\tau})~.$ 

\pagebreak

Theorem 2~([6]). ~Let $(\widehat{A},\mathbb{R}^n,\beta)$ be a $F^*$-dynamical 
system with a faithful $\beta$-invariant continuous trace $\tau$,and $\Xi$ a 
finitely generated projective right $\widehat{A}$-module. If $\beta$ commutes 
with $\overline{\alpha}$, then there exist a $F^*$-dynamical system $(A,\mathbb{R}^n,\beta_A)$ with a faithful $\beta_A$-invariant continuous trace 
$\tau_A$, and a finitely generated projective right $A$-module $\Xi_A$ such that \[ \mathcal{M}^{(\widehat{A},\mathbb{R}^n,\beta,\tau)}(\Xi) \approx \mathcal{M}^{(A,\mathbb{R}^n,\beta_A,\tau_A)}(\Xi_A)  ~.\]

We want to apply the above theorem to the following important example which 
appears as a Higgs branch of the theory of D0-branes bound to D4-branes by the 
expectation value of the B-field as well as a regularized version of the target  space of supersymmetric quantum mechanics arising in the light cone description of (2,0) superconformal theories in six dimensions, although its algebraic 
structure has already been established in the example 10.1 of [5](cf.[3],[4]):\\

 \quad Let $\mathbb{R}^4_{\theta}$ be the noncommutative $\mathbb{R}^4$ for an 
antisymmetric 4x4 matrix $\theta = (\theta_{i,j})$, in other words, $\mathbb{R}^4_{\theta}$ is the $F^*$-algebra generated by 4-self adojoint elements $\{x_i\}_{i=1}^4$  with the property that \\

\noindent
$(1)~~~~~~~~~~~~~~~~~ [x_i,x_j] = \theta_{i,j}     $ \\

\noindent
$(i,j=1,\cdots,4)$. In other words, 
\[ \mathbb{R}^4_{\theta}~=~\{ \sum_{i_1,i_2,i_3,i_4 \in \mathbb{N}} c_{i_1,i_2,i_3,i_4}~x_1^{i_1}x_2^{i_2}x_3^{i_3}x_4^{i_4}~|~c \in S(\mathbb{N})~\}  \]
\noindent
where $S(\mathbb{N})$ is the set of all rapidly decreasing complex valued 
functions on $\mathbb{N}$. Let $x^i=\theta^{i,j}x_j~(i,j=1,\cdots,4)$ 
where $(\theta^{i,j})$ is the inverse matrix of 
$(\theta_{i,j})$. Then the $F^*$-algebra $\mathbb{R}^4_{\theta}$ 
depends essentially on one positive real number denoted by the same symbol 
$\theta$, which satisfy the following relation: \\

\noindent
$(2) ~~ [z_i^*,z_i] = \theta~,~[z_i,z_j]=[z_i^*,z_j]=0 ~~(i,j=0,1,i\neq j) $ \\

\noindent
where $z_0=x^1+\sqrt{-1}x^2,z_1=x^3+\sqrt{-1}x^4$ and $z_i^*$ are the conjugate operators of $z_i$. Let us consider the canonical action $\alpha$ of 
$\mathbb{R}^4$ on $\mathbb{R}^4_{\theta}$ defined by \\

\noindent
$(3)~~~~~~~~~~~~~~~~~ \alpha_{t_i}(x_i) = x_i + t_i     $ \\

\noindent
$(t_i \in \mathbb{R},i=1,\cdots,4)$. Then it is easily seen that the triplet 
$(\mathbb{R}^4_{\theta},\mathbb{R}^4,\alpha)$ is a $F^*$-dynamical system, and 
we easily see that \\

\noindent
$(4)~~~~~~~~~~~~~~~~~ \alpha_{w_i}(z_i)= z_i + w_i     $ \\

\noindent
$(w_i \in \mathbb{C}, i=0,1)$. 
By $(2)$, $\mathbb{R}^4_{\theta}$ is nothing but the $F^*$-tensor product $A_0 
\otimes A_1$ where $A_i$ are the $F^*$-algebras generated by $z_i~(i=0,1)$. 
We now check the algebraic structure of $A_i$. By $(2)$, it follows from [3](cf.[4]) that there exist two Fock spaces $H_i$ such that \\

$~~~~~~z_i(\xi^i_n)=\sqrt{(n+1)\theta}~\xi^0_{n+1}~,~z_i^*(\xi^i_n)=\sqrt{n \theta}~\xi^i_{n-1}$, \\

\noindent
where $\{\xi^i_n\}$ are complete orthonormal systems of $H_i$ 
with respect to the following inner product:
\[ <~f~|~g~>=\sum (n+1)\theta~f(n)\overline{g(n)} \]
\noindent
for two $\mathbb{C}$-valued functions $f,g$ on $\mathbb{N}$ such that  
\[ \sum (n+1)\theta~|f(n)|^2 < \infty~,~\sum (n+1)\theta~|g(n)|^2 < \infty ~.\]
\noindent
for $i=0,1$. 
We may assume that the $A_i$ act on $H_i$ irreducibly. Then it also follows 
from [4] that the $F^*$-algebras $A_i$ are isomorphic to the $F^*$-algebras 
$\mathcal{K}^{\infty}(H_i)$ defined by
\[  \mathcal{K}^{\infty}(H_i)=\{T \in \mathcal{K}(H_i)~|~\{\lambda_k\} \in 
S(\mathbb{N}) \}  \]
\noindent
where $\{\lambda_k\}$ are all eigen values of $T$ and $S(\mathbb{N})$ are the 
set of all sequences $\{c_n\}$ of $\mathbb{C}$ with 
$sup_{n\geq 1}~(1+|n|)^k|c_n| < \infty $ for all $k \geq 0$. Therefore, 
the $F^*$-algebra $\mathbb{R}^4_{\theta}$ is isomorphic to $\mathcal{K}^{\infty}(H_0 \otimes H_1)$. We then have the following proposition: \\

Proposition 3~(cf:[5]). \quad If $\theta \neq 0$, then $\mathbb{R}^4_{\theta}$ is isomorphic to $\mathcal{K}^{\infty}(L^2(\mathbb{C}^2))$ as a $F^*$-algebra. \\

\noindent
By the above Proposition, $\mathcal{K}^{\infty}(L^2(\mathbb{C}^2))$ is the $F^*$-crossed product $S(\mathbb{C}^2) \rtimes_{\tau} \mathbb{C}^2$ of $S(\mathbb{C}^2)$ by the shift action $\tau$ of $\mathbb{C}^2$. We then consider the action 
$\alpha$ defined before. By $(4)$, it follows from [R] that $\alpha$ plays a 
role of the dual action of $\tau$. Then the $F^*$-crossed product $\widehat{\mathbb{R}^4_{\theta}}$ of $\mathbb{R}^4_{\theta}$ by the action $\alpha$ of $\mathbb{R}^4$ is isomorphic to the $F^*$-crossed product $\mathcal{K}^{\infty}(L^2(\mathbb{C}^2)) \rtimes_{\widehat{\tau}} \mathbb{C}^2$, where $\widehat{\tau}$ is 
the dual action of $\tau$. Then it is isomorphic to $S(\mathbb{C}^2) \otimes 
\mathcal{K}^{\infty}(L^2(\mathbb{C}^2))$ as a $F^*$-algebra. \\
~~~~ We now consider a finitely generated projective right $\mathbb{R}^4_{\theta}$-module $\Xi$. Then there exist an integer $n \geq 1$ and a projection $P \in  M_n(\mathcal{M}(\mathbb{R}^4_{\theta}))$ such that $\Xi=P((\mathbb{R}^4_{\theta})^n)$. where $\mathcal{M}(\mathbb{R}^4_{\theta})$ is the $F^*$-algebra consisting of all bounded linear operators $T$ on $L^2(\mathbb{C}^2)$ whose kernel functions $T(\cdot~,~\cdot)$ are $\mathbb{C}$-valued bounded $C^{\infty}$-functions of $\mathbb{C}^2 \times \mathbb{C}^2$. 
Let us take the canonical faithful trace $Tr$ on $\mathbb{R}^4_{\theta}$ because of Proposition 1. Then we consider the moduli space: \\

\noindent
$\mathcal{M}^{(\mathcal{K}^{\infty}(L^2(\mathbb{C}^2)),\mathbb{C}^2,\alpha,Tr)}(\Xi)$ of $\Xi$ for $(\mathcal{K}^{\infty}(L^2(\mathbb{C}^2)),\mathbb{C}^2,\alpha,Tr)$. \\

We want to describe $P$ cited above as a precise fashion. Actually, we know that \[\mathcal{K}^{\infty}(L^2(\mathbb{C}^2)) \cong S(\mathbb{C}^2) \rtimes_{\lambda} \mathbb{C}^2 \]
\noindent
where $\cong$ means isomorphism as a $F^*$-algebra. $\lambda$ is the shift 
action of $\mathbb{C}^2$ on $S(\mathbb{C}^2)$. 
Then it follows that
\[ M_n(\mathcal{K}^{\infty}(L^2(\mathbb{C}^2))) \cong M_n(S(\mathbb{C}^2)) 
\rtimes_{\lambda^n} \mathbb{C}^2 \]
\noindent
where 
\[ \lambda^n_{w}(f)(w')=f(w'-w) \]
\noindent
for $f \in M_n(S(\mathbb{C}^2)),w,w' \in \mathbb{C}^2$. Let $\overline{\lambda^n}$ be the action of $\mathbb{C}^2$ on $M_n(\mathcal{K}^{\infty}(L^2(\mathbb{C}^2)))$ associated with $\lambda^n$ satisfying Theorem 2. 
It follows from  Proposition 3 that \\

$\mathcal{M}^{(\mathbb{R}^4_{\theta},\mathbb{R}^4,\alpha,Tr)}(\Xi) \approx \mathcal{M}^{(\mathcal{K}^{\infty}(L^2(\mathbb{C}^2),\mathbb{C}^2,\alpha,Tr)}(\Xi_1)$ \\

$~~~~~~~~~~~~~~~~~~~~ \approx \mathcal{M}^{(S(\mathbb{C}^2) \rtimes_{\lambda} \mathbb{C}^2,\mathbb{C}^2,\widehat{\lambda},\widehat{\int_{\mathbb{C}^2}dz})}(\Xi_2)~,$ \\
\noindent
where 
\[ \Xi_1=P_1(\mathcal{K}^{\infty}(L^2(\mathbb{C}^2)^n)~,~\Xi_2=P_2((S(\mathbb{C}^2) \rtimes_{\lambda} \mathbb{C}^2)^n) \]
\noindent
for the two projections $P_j~(j=1,2)$ with the property that 
\[ P_1 \in M_n(\mathcal{M}(\mathcal{K}^{\infty}(L^2(\mathbb{C}^2)))~,~P_2 \in 
M_n(\mathcal{M}(S(\mathbb{C}^2)) \rtimes_{\lambda} \mathbb{C}^2) \]
\noindent
corresponding to $\Xi$, where $\mathcal{M}(S(\mathbb{C}^2)$ is the $F^*$-algebra consisting of all $\mathbb{C}$-valued bouded $C^{\infty}$-functions on $\mathbb{C}^2$ and $\lambda$ is the shift action of $\mathbb{C}^2$ on $\mathcal{M}(S(\mathbb{C}^2))$. By its definition, we know that
\[\overline{\lambda}_w=\widehat{\lambda}_w \circ \widetilde{\lambda}_w ~,~(w \in \mathbb{C}^2) \]
\noindent
where $\widehat{\lambda}$ is the dual action of $\lambda$ and 
\[\widetilde{\lambda}_w(x)(w')=\lambda_wx(w') \]
\noindent
for all $x \in S(\mathbb{C}^2) \rtimes_{\lambda} \mathbb{C}^2$ and $w,w' \in \mathbb{C}^2$. Hence $\widehat{\lambda}$ commutes with $\overline{\lambda}$, which implies by Theorem 2 that there exist a $F^*$-dynamical system $(S(\mathbb{C}^2),\mathbb{C}^2,\widehat{\lambda}_{S(\mathbb{C}^2)},\int_{\mathbb{C}^2}dz)$ and a finitely generated projective right $A$-module $(\Xi_2)_{S(\mathbb{C}^2)}$ such that \\

\noindent
$ ~~~~~~~\mathcal{M}^{(S(\mathbb{C}^2) \rtimes_{\lambda} \mathbb{C}^2,\mathbb{C}^2,\widehat{\lambda},\widehat{\int_{\mathbb{C}^2}dz})}(\Xi_2)$ \\

$~~~~~~~~~~~\approx ~~~\mathcal{M}^{(S(\mathbb{C}^2),\mathbb{C}^2,\widehat{\lambda}_{S(\mathbb{C}^2)},\int_{\mathbb{C}^2}dz)}((\Xi_2)_{S(\mathbb{C}^2)}) ~.$ \\

\noindent
We know that there exist an integer $m \geq 1$ and a projection $Q \in M_m(\mathcal{M}(S(\mathbb{C}^2)))$ such that
\[ (\Xi_2)_{S(\mathbb{C}^2)}=Q(S(\mathbb{C}^2)^m)~.\]
\noindent
Moreover, it follows from the definition that the action $\widehat{\lambda}_{S(\mathbb{C}^2)}$ is nothing but $\lambda$. We now determine the moduli space $\mathcal{M}^{(S(\mathbb{C}^2),\mathbb{C}^2,\lambda,\int_{\mathbb{C}^2}dz)})(Q(S(\mathbb{C}^2)^m)$ in what follows: Since $Q \in M_m(\mathcal{M}(S(\mathbb{C}^2)))$ and 
\[ M_m(S(\mathbb{C}^2)) \cong S(\mathbb{C}^2,M_m(\mathbb{C})) ~,\]
\noindent
then it also follows from Theorem 2 that there exist a finitely generated projective right $\mathbb{C}$-module $Q(S(\mathbb{C}^2)^m)_{\mathbb{C}}$ such that \\

$~~~~~~~~\mathcal{M}^{(S(\mathbb{C}^2),\mathbb{C}^2,\widehat{\lambda}_{S(\mathbb{C}^2)},\int_{\mathbb{C}^2}dz)}(Q(S(\mathbb{C}^2)^m)) ~$  \\

$~~~~~~~~~~~\approx ~~~\mathcal{M}^{(\mathbb{C},\mathbb{C}^2,\lambda_{\mathbb{C}},1)}(Q(S(\mathbb{C}^2)^m)_{\mathbb{C}}) ~.$ \\

\noindent
Since $Q(S(\mathbb{C}^2)^m)_{\mathbb{C}}$ is a finitely generated projective right $\mathbb{C}$-module, then its construction tells us that there exists a projection $R \in M_m(\mathbb{C})$ such that 
\[  Q(S(\mathbb{C}^2)^m)_{\mathbb{C}})~=~R(\mathbb{C}^m) ~.\]
\noindent
Summing up the argument discussed above, we deduce that
\[ \mathcal{M}^{(\mathbb{R}^4_{\theta},\mathbb{R}^4,\alpha,Tr)}(\Xi) \approx \mathcal{M}^{(\mathbb{C},\mathbb{C}^2,\iota,1)}(R(\mathbb{C}^m))~.\]
\noindent
By the definition of the moduli space, we deduce that \\

$~~~~\mathcal{M}^{(\mathbb{C},\mathbb{C}^2,\iota,1)}(R(\mathbb{C}^m))$ \\

$~~~~~~~~~~~~~~\approx \mathrm{End}_{\mathbb{C}}(R(\mathbb{C}^m))_{sk}/U(\mathrm{End}_{\mathbb{C}}(R(\mathbb{C}^m))~,$ \\

\noindent
where 
\[ \mathrm{End}_{\mathbb{C}}(R(\mathbb{C}^m))_{sk}~~~(\mathrm{resp}.~U(\mathrm{End}_{\mathbb{C}}(R(\mathbb{C}^m))) \]
\noindent
is the set of all skew adjoint (resp.~unitary) elements in 
$\mathrm{End}_{\mathbb{C}}(R(\mathbb{C}^m))~.$ 
Since $\mathrm{End}_{\mathbb{C}}(R(\mathbb{C}^m))=M_k(\mathbb{C})$ for some 
natural number $k~(m \geq k)$, it follows by using diagonalization that
\[\mathrm{End}_{\mathbb{C}}(R(\mathbb{C}^m))_{sk}/U(\mathrm{End}_{\mathbb{C}}(R(\mathbb{C}^m)) \approx \mathbb{R}^k ~,\]
\noindent
which implies the following theorem: \\

Theorem 4. \quad Let $\mathbb{R}^4_{\theta}$ be the deformation quantization of   $\mathbb{R}^4$ with respect to a skew symmetric matrix $\theta$ and take the  $F^*$-dynamical system $(\mathbb{R}^4_{\theta},\mathbb{R}^4,\alpha)$ with a 
canonical faithful $\alpha$-invariant trace $Tr$ of $\mathbb{R}^4_{\theta}$, 
where $\alpha$ is the translation action of $\mathbb{R}^4$ on $\mathbb{R}^4_{\theta}$. Suppose $\Xi$ is a finitely generated projective right $\mathbb{R}^4_{\theta}$-module, then there exists a natural number $k$ such that 
\[\mathcal{M}^{(\mathbb{R}^4_{\theta},\mathbb{R}^4,\alpha,Tr)}(\Xi) \approx 
\mathbb{R}^k  ~.\]

Remark. \quad The above theorem only states the topological data of the moduli 
spaces of Yang-Mills connections. We would study their both differential and 
holomorphic structures in a forthcoming paper (cf:[2]). 

\pagebreak
\begin{center}

References

\end{center}
\vspace{5mm}

\noindent
[1]~K.Furuuchi:~Instantons on Noncommutative $\mathbb{R}^4$ and \\
~~~~Projection Operators. arXiv:hep-th/9912047. \\

\noindent
[2]~H.Nakajima:~Resolutions of Moduli Spaces of Ideal \\
~~~~Instantons on $\mathbb{R}^4$.~World Scientific.129-136(1994).\\

\noindent
[3]~N.Nekrasov and A.Schwarz:~Instantons on Noncom-\\
~~~~mutative $\mathbb{R}^4$,and $(2,0)$ Superconformal Six Dimen-\\
~~~~~sional Theory.~Commun.Math.Phy.198,689-703(1998),\\

\noindent
[4]~C.R.Putnam:~Commutation Properties of Hilbert Space \\
~~~~~Operators and Related Topics.~Springer-Verlag (1967).\\

\noindent
[5]~M.A.Rieffel:~Deformation Quantization for Actions of \\
 ~~~~$\mathbb{R}^d$.Memoires AMS.506(1993).\\

\noindent
[6]~H.Takai:~Yang-Mills Theory for Noncommutative Flows. \\
~~~~arXiv:math-ph/040326.

\end{document}